\documentclass[conference]{IEEEtran}
\IEEEoverridecommandlockouts
\usepackage{cite}
\usepackage{amsmath,amssymb,amsfonts}
\usepackage{algorithmic}
\usepackage{graphicx}
\usepackage{textcomp}
\usepackage{xcolor}
\def\BibTeX{{\rm B\kern-.05em{\sc i\kern-.025em b}\kern-.08em
    T\kern-.1667em\lower.7ex\hbox{E}\kern-.125emX}}
\begin{document}

\title{Reproducing the acoustic velocity vectors in a circular listening area\\
\thanks{Frank Jiarui Wang was supported by The Australian National University's Fee Remission Scholarship.}
}

\author{\IEEEauthorblockN{Frank Jiarui Wang}
\IEEEauthorblockA{
\textit{The Australian National University}\\
Canberra, Australia \\
u5879960@anu.edu.au}
\and
\IEEEauthorblockN{Jihui Aimee Zhang}
\IEEEauthorblockA{
\textit{University of Southampton}\\
Southampton, UK \\
Aimee\_Jihui.Zhang@soton.ac.uk}
\and
\IEEEauthorblockN{Thushara Abhayapala, Prasanga Samarasinghe}
\IEEEauthorblockA{
\textit{The Australian National University}\\
Canberra, Australia \\
thushara.abhayapala@anu.edu.au}

}

\maketitle

\begin{abstract}
Acoustic velocity vectors are important for human's localization of sound at low frequencies. This paper proposes a sound field reproduction algorithm, which matches the acoustic velocity vectors in a circular listening area. In previous work, acoustic velocity vectors are matched either at sweet spots or on the boundary of the listening area. Methods based on sweet spots experience performance degradation when the listener moves away from sweet spots, whereas measuring the acoustic velocity vectors on the boundary requires complicated measurement setup. This paper proposes the radial independent cylindrical harmonic coefficients of the acoustic velocity vectors (CHV-indR coefficients) in the circular listening area, which are calculated from the cylindrical harmonic coefficients of the pressure in the circular listening area by using the sound field translation formula. The cylindrical harmonic coefficients of the pressure can be measured by a circular microphone array, which can be bought off-the-shelf. By matching the CHV-indR coefficients, the acoustic velocity vectors are reproduced throughout the listening area. Simulations show that at low frequencies, where the acoustic velocity vectors are the dominant factor for localization, the proposed reproduction method based on matching the CHV-indR coefficients results in higher accuracy in reproduced acoustic velocity vectors when compared with traditional method based on matching the cylindrical harmonic coefficients of the pressure.
\end{abstract}


\section{Introduction}
Acoustic velocity vectors are the dominant factor for human's localization of sound at low frequencies, especially below 700 Hz \cite{Gerzon1992, gerzon1992G}, due to their relevance to the interaural time difference. Acoustic velocity vectors are commonly used in Ambisonic encoding and decoding  \cite{Arteaga2013, AmbisonicsBook}. For spatial sound field reproduction system, whose aim is to reproduce a sound field over a region or an area, incorporating acoustic velocity vectors can result in improved perception.

There are multiple ways to incorporate acoustic velocity vectors in spatial sound field reproduction system. First, it is possible to control the acoustic velocity vectors at multiple sweet spots \cite{Wen2021, Wen2023}, which can cause a reduction in performance when listener moves beyond the sweet spots. Second, based on the Kirchhoff-Helmholtz integral \cite{SporsReview}, reproduction can be achieved by jointly reproducing the pressure and the acoustic velocity vectors on the boundary of the listening region \cite{Shin2016, JPVM, JPVMPlus}. Measuring the acoustic velocity vectors at multiple control points on the boundary requires complicated and potentially expensive setup, which may not be feasible due to space and financial constraints. Recently, in \cite{ZuoVel}, the spherical harmonic coefficients of the acoustic velocity vectors (SHV coefficients) on the spherical listening region's boundary were derived from the spherical harmonic coefficients of the pressure within the spherical listening region. Therefore, the SHV coefficients can be derived from spherical microphone array measurements \cite{Thushara2002, Meyer2002}. Using a spherical microphone array is easier and possibly cheaper than using multiple velocity probes. Third, the acoustic velocity vectors can be reproduced throughout the listening region, i.e., instead of at sweet spots or on the boundary, the acoustic velocity vectors are reproduced at \textbf{every} point inside the listening region. 

A spherical region can be modeled as the superposition of concentric spheres whose radii are infinitesimally close to each other. The SHV coefficients in \cite{ZuoVel} have a radial component, i.e., they need to be calculated for each concentric sphere. It is possible to match the SHV coefficients on multiple concentric spheres to achieve reproduction of the acoustic velocity vectors in the spherical listening region. However, the reproduction performance depends on radial sampling, i.e., the selection of the radii of the concentric spheres. In \cite{FrankSPL}, the SHV coefficients independent of the radial distance (SHV-indR) were proposed. The SHV-indR coefficients can be derived from spherical microphone array measurements of the pressure within the spherical listening region by using the sound field translation formula. To characterize the acoustic velocity vectors within the spherical listening region, one set of SHV-indR coefficients is sufficient due to the radial independence. The SHV-indR coefficients were used in spatial sound field reproduction and demonstrated better performance at low frequencies when compared with method that matches the spherical harmonic coefficients of the pressure \cite{FrankSPL}. 

In some circumstances, it might be sufficient to reproduce the sound field on a two-dimensional (2D) plane, e.g., the plane where the listener's head is located. 2D sound field are assumed to be height-invariant, i.e., the changes in physical quantities in the $z$-axis (vertical) direction are ignored. A 2D or height-invariant sound field can be represented using cylindrical harmonic expansion \cite{WilliamsCyl}. This paper builds on \cite{FrankSPL} and proposes the radial independent cylindrical harmonic coefficients of the acoustic velocity vectors (CHV-indR coefficients) in a circular listening area. Like the SHV-indR coefficients, the CHV-indR coefficients are independent of the radial distance. By using the sound field translation formula, the CHV-indR coefficients can be derived from the cylindrical harmonic coefficients of the pressure inside the circular listening area. The cylindrical harmonic coefficients of the pressure can be measured by a circular microphone array. This paper also shows that the CHV-indR coefficients can be used in the reproduction of 2D or height-invariant sound field. Moreover, reproduction based on matching the CHV-indR coefficients achieves lower direction error in the reproduced acoustic velocity vectors when compared with method that matches the cylindrical harmonic coefficients of the pressure. 

\section{Acoustic velocity vectors in a circular area}
\subsection{The geometric model}
\begin{figure}[t]
  \centering
  \includegraphics[trim = 40mm 25mm 40mm 25mm, clip, width = 0.7\columnwidth]{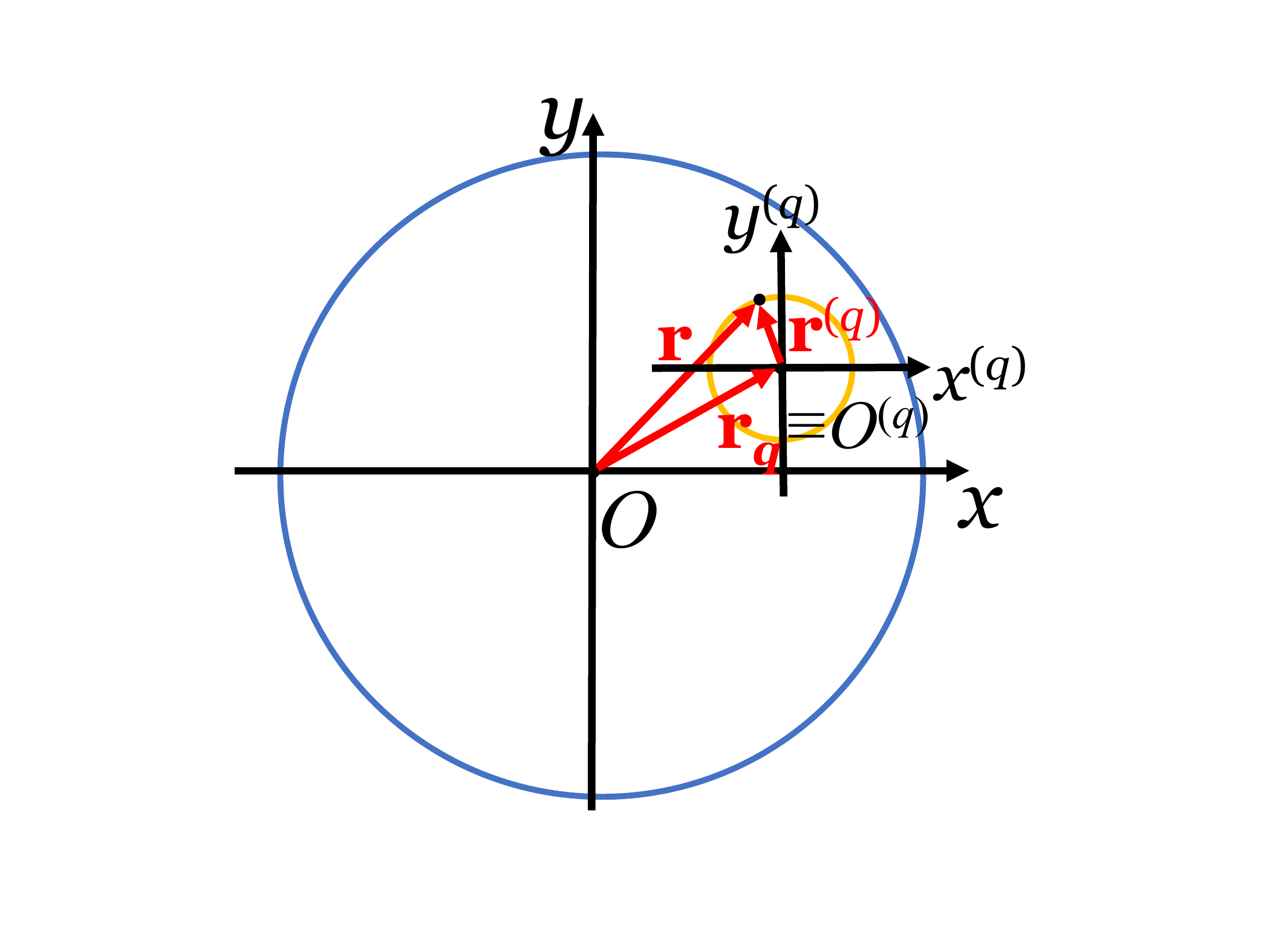}
  \vskip -3mm
  \caption{Setup of the geometric model. The listening region is bounded by the blue circle. $\mathbf{r}_{\boldsymbol{q}}$ is a point in the listening region.}
  \vskip -3mm
  \label{Fig:model_setup}
  \vskip -2mm
\end{figure}

Figure \ref{Fig:model_setup} shows the setup of the geometric model. This section aims to find the CHV-indR coefficients, which represent the acoustic velocity vectors within the listening area bounded by the blue circle. The derivation first finds the acoustic velocity vector at point $\mathbf{r_{\boldsymbol{q}}}$. The local $x^{(q)}y^{(q)}z^{(q)}$ coordinate system is the translation of the global $xyz$ coordinate system with $\mathbf{r_{\boldsymbol{q}}}$ as the translation vector. In this paper, the superscript indicates the coordinate system used for expressing the location. When there are no superscripts, the location is expressed using the $xyz$ coordinate system. For point $\mathbf{r}$, the relationship $\mathbf{r} = \mathbf{r_{\boldsymbol{q}}} + \mathbf{r}^{(q)}$ holds. 

\section{Acoustic velocity vector at a point}
Consider the local region in Figure \ref{Fig:model_setup} bounded by the yellow circle. Assume the sound field is height invariant, by using the cylindrical harmonic expansion \cite{WilliamsCyl}, the pressure at $\mathbf{r}^{(q)} \equiv (r^{(q)}, \phi^{(q)})$ is 
\begin{equation}
\label{Eq:pressure_local}
    p(k, \mathbf{r}^{(q)}) = \sum_{m = -M}^{M} \alpha_{m}(k, \mathbf{r}_{\boldsymbol{q}}) J_{m}(kr^{(q)}) e^{im\phi^{(q)}}
\end{equation}
in which $k$ is the wavenumber, $\alpha_{m}(k, \mathbf{r}_{\boldsymbol{q}})$ are the cylindrical harmonic coefficients of the local pressure, $J_{m}(\cdot)$ is the Bessel function of the first kind and $M$ is the truncation order. The radial derivative of the Bessel function at the origin satisfies
\begin{equation}
    \frac{\partial J_{m}(kr^{(q)})}{\partial r^{(q)}} \bigg\vert_{r^{(q)} = 0} = \frac{1}{2} k \delta_{m, 1} - \frac{1}{2} k \delta_{m, -1} 
\end{equation}
in which $\delta_{m, \pm1}$ are Kronecker delta functions. The acoustic velocity vector at $\mathbf{r}_{\boldsymbol{q}}$ is
\begin{align}
    &V_{\boldsymbol{\hat{x}}}(k, \mathbf{r}_{\boldsymbol{q}}) = \frac{i}{k\rho_{0}c}  \frac{\partial p(k, \mathbf{r}^{(q)})}{\partial x} \bigg\vert_{r^{(q)} = 0} \nonumber\\
    &= \frac{i}{k\rho_{0}c} \sum_{m = -M}^{M} \alpha_{m}(k, \mathbf{r}_{\boldsymbol{q}}) \frac{\partial J_{m}(kr^{(q)})}{\partial r^{(q)}} \bigg\vert_{r^{(q)} = 0}  e^{im0} \nonumber\\
    &= \frac{1}{2}\frac{i}{\rho_{0}c} \big[\alpha_{1}(k, \mathbf{r}_{\boldsymbol{q}}) - \alpha_{-1}(k, \mathbf{r}_{\boldsymbol{q}})\big] \label{Eq:Vx_local}\\
    &V_{\boldsymbol{\hat{y}}}(k, \mathbf{r}_{\boldsymbol{q}}) = \frac{i}{k\rho_{0}c}  \frac{\partial p(k, \mathbf{r}^{(q)})}{\partial y} \bigg\vert_{r^{(q)} = 0} \nonumber\\
    &= \frac{i}{k\rho_{0}c} \sum_{m = -M}^{M} \alpha_{m}(k, \mathbf{r}_{\boldsymbol{q}}) \frac{\partial J_{m}(kr^{(q)})}{\partial r^{(q)}} \bigg\vert_{r^{(q)} = 0}  e^{im\frac{\pi}{2}} \nonumber\\
    &= -\frac{1}{2}\frac{1}{\rho_{0}c} \big[\alpha_{1}(k, \mathbf{r}_{\boldsymbol{q}}) + \alpha_{-1}(k, \mathbf{r}_{\boldsymbol{q}})\big]  \label{Eq:Vy_local}
\end{align}
\section{Acoustic velocity vector in a circular area}
Using the global $xyz$ coordinate system, the pressure at point $\mathbf{r} \equiv (r, \phi)$ is
\begin{equation}
\label{Eq:pressure_global}
     p(k, \mathbf{r}) = \sum_{\nu = -V}^{V} \beta_{\nu}(k) J_{\nu}(kr) e^{i\nu\phi}
\end{equation}
in which $V$ is the truncation order and $\beta_{\nu}(k)$ are the cylindrical harmonic coefficients of the pressure inside the listening area. Substituting the sound field translation formula \cite{PrasangaThesis}
\begin{align}
    &J_{\nu}(kr) e^{i\nu\phi} \nonumber\\ &= \sum_{m=-\infty}^{\infty} J_{m-\nu}(kr_{q}) e^{i(\nu-m)(\pi+\phi_{q})} J_{m}(kr^{(q)}) e^{im\phi^{(q)}},
\end{align}
into \eqref{Eq:pressure_global}, the pressure  
\begin{align}
    &p(k, \mathbf{r}^{(q)}) \nonumber\\
    &= \sum_{m=-M}^{M} \underbrace{\bigg[\sum_{\nu = -V}^{V} \beta_{\nu}(k) J_{m-\nu}(kr_{q}) e^{i(\nu-m)(\pi+\phi_{q})}\bigg]}_{\alpha_{m}(k, \mathbf{r}_{\boldsymbol{q}})} \nonumber\\ &\qquad J_{m}(kr^{(q)}) e^{im\phi^{(q)}}.
\end{align}
Let $n = \nu-m$,
\begin{equation}
\label{Eq:CH_decomp_local_CH}
    \alpha_{m}(k, \mathbf{r}_{\boldsymbol{q}}) = \sum_{n = -V-m}^{V-m} \beta_{n+m}(k) J_{-n}(kr_{q}) e^{in(\pi+\phi_{q})}.
\end{equation}
Using the property \cite{BesselNegOrder}
\begin{equation}
    J_{-n}(\cdot) = (-1)^{n} J_{n}(\cdot),
\end{equation}
equation \eqref{Eq:CH_decomp_local_CH} becomes
\begin{equation}
\label{Eq:CH_decomp_local_CH_final}
    \alpha_{m}(k, \mathbf{r}_{\boldsymbol{q}}) = \sum_{n = -V-m}^{V-m} \underbrace{\beta_{n+m}(k)}_{\gamma_{n}^{(m)}(k)} J_{n}(kr_{q}) e^{in\phi_{q}}
\end{equation}
in which $\gamma_{n}^{(m)}$ are the cylindrical harmonic coefficients of $\alpha_{m}(k, \mathbf{r}_{\boldsymbol{q}})$.

To find the acoustic velocity vector at $\mathbf{r}_{\boldsymbol{q}}$, it is only necessary to calculate $\alpha_{\pm 1}(k, \mathbf{r}_{\boldsymbol{q}})$, i.e., $m = \pm 1$. To comply with the general form of cylindrical harmonic expansion where the upper summation limit and the lower summation limit are symmetric, (i.e., the summation is of the form $\sum_{n = -N'}^{N'}$), the summation in \eqref{Eq:CH_decomp_local_CH_final} is truncated to $\sum_{n = -V+1}^{V-1}$. From \eqref{Eq:CH_decomp_local_CH_final}, the mapping from $\beta_{\nu}(k)$ to $\gamma_{n}^{(\pm 1)}(k)$ can be expressed in matrix form
\begin{align}
    \pmb{\gamma}^{(1)}(k) &= A^{(1)} \pmb{\beta}(k) \\
    \pmb{\gamma}^{(-1)}(k) &= A^{(-1)} \pmb{\beta}(k) 
\end{align}
with
\begin{align}
    A^{(1)} &= \delta_{n, \nu-1} \label{Eq:CHP2CH1}\\
    A^{(-1)} &= \delta_{n, \nu+1}. \label{Eq:CHP2CH_neg1}
\end{align}
The column vectors $\pmb{\beta}(k) = [\beta_{-V}(k), \cdots, \beta_{V}(k)]^{T}$ and $\pmb{\gamma}^{(\pm 1)}(k) = [\gamma_{-V+1}^{(\pm 1)}(k), \cdots, \gamma_{V-1}^{(\pm 1)}(k), ]^{T}$. The matrices $A^{(1)}$ and $A^{(-1)}$ are of dimension $(2V-1)$-by-$(2V+1)$. As \eqref{Eq:CHP2CH1} and \eqref{Eq:CHP2CH_neg1} suggest, $A^{(1)}$ and $A^{(-1)}$ are independent of the wavenumber $k$. 

Since the acoustic velocity vectors are the linear combinations of $\alpha_{1}(k, \mathbf{r}_{\boldsymbol{q}})$ and $\alpha_{-1}(k, \mathbf{r}_{\boldsymbol{q}})$, it is possible to express them using cylindrical harmonic expansion as
\begin{equation}
    V_{\boldsymbol{\hat{e}}}(k,\mathbf{r}_{\boldsymbol{q}}) = \sum_{n = -V+1}^{V-1} (\zeta_{\boldsymbol{\hat{e}}})_{n}(k) J_{n}(kr_{q}) e^{in\phi_{q}}
\end{equation}
with $\boldsymbol{\hat{e}} \in \{\boldsymbol{\hat{x}}, \boldsymbol{\hat{y}}\}$. $(\zeta_{\boldsymbol{\hat{e}}})_{n}(k)$ are the CHV-indR coefficients. It is clearly shown that $(\zeta_{\boldsymbol{\hat{e}}})_{n}(k)$ are independent of the radial distance $r_{q}$. Using \eqref{Eq:Vx_local}, \eqref{Eq:Vy_local} and the expansion in \eqref{Eq:CH_decomp_local_CH_final}, 
\begin{align}
    (\zeta_{\boldsymbol{\hat{x}}})_{n}(k) &= \frac{1}{2} \frac{i}{\rho_{0}c} [\gamma_{n}^{(1)}(k) - \gamma_{n}^{(-1)}(k)] \\
     (\zeta_{\boldsymbol{\hat{y}}})_{n}(k) &= -\frac{1}{2} \frac{1}{\rho_{0}c} [\gamma_{n}^{(1)}(k) + \gamma_{n}^{(-1)}(k)].
\end{align}
The mapping from $\beta_{\nu}(k)$ to $ (\zeta_{\boldsymbol{\hat{e}}})_{n}(k)$ can also be expressed using matrix form
\begin{equation}
\label{Eq:CHP_2_CHV}
     \boldsymbol{\zeta}_{\boldsymbol{\hat{e}}}(k) = A_{\boldsymbol{\hat{e}}} \,\pmb{\beta}(k),
\end{equation}
where 
\begin{align}
    A_{\boldsymbol{\hat{x}}} &= \frac{1}{2}\frac{i}{\rho_{0}c} [A^{(1)} - A^{(-1)}]\\
    A_{\boldsymbol{\hat{y}}} &= -\frac{1}{2}\frac{1}{\rho_{0}c} [A^{(1)} + A^{(-1)}].
\end{align}
The matrices $A_{\boldsymbol{\hat{e}}}$ are of dimension $(2V-1)$-by-$(2V+1)$. If the cylindrical harmonic coefficients $\beta_{\nu}(k)$ of the pressure are measured up to $\nu = \pm V$, the CHV-indR coefficients $(\zeta_{\boldsymbol{\hat{e}}})_{n}(k)$ can be calculated up to order $n = \pm (V-1)$. The column vectors $\boldsymbol{\zeta}_{\boldsymbol{\hat{e}}}(k) = [(\zeta_{\boldsymbol{\hat{e}}})_{-V+1}(k), \cdots, (\zeta_{\boldsymbol{\hat{e}}})_{V-1}(k)]^{T}$. Moreover, the matrices $A_{\boldsymbol{\hat{e}}}$ are independent of the wavenumber $k$.

\section{Illustration of acoustic velocity vectors in a circular area}
This section illustrates the acoustic velocity vectors in a circular area when the source is a plane wave and a line source. When the source is a plane wave with incident direction $\phi_{\text{pw}}$, using Jacobi-Anger expansion, the cylindrical harmonic coefficients of the pressure in \eqref{Eq:pressure_global} are \cite{MRPT2014}
\begin{equation}
    \beta_{\nu}(k) = i^{\nu} e^{-i\nu\phi_{\text{pw}}}.
\end{equation}
When the source is an infinite line source in $z$ direction located at $\mathbf{r_{s}} \equiv (r_{s}, \phi_{s})$ on the two-dimensional $xy$ plane, the pressure at $\mathbf{r} \equiv (r, \phi, z = 0) \equiv (r, \phi)$ can be expressed using \eqref{Eq:pressure_global}, where \cite{Poletti2010}
\begin{equation}
    \beta_{\nu}(k) = \frac{-i}{4} H_{\nu}^{(2)}(kr_{s}) e^{-i\nu\phi_{s}}
\end{equation}
in which $H_{\nu}^{(2)}(\cdot)$ is the Hankel function of the second kind.

\begin{figure}[t]
  \centering
  \includegraphics[trim = 15mm 30mm 25mm 45mm, clip, width = 0.9\columnwidth]{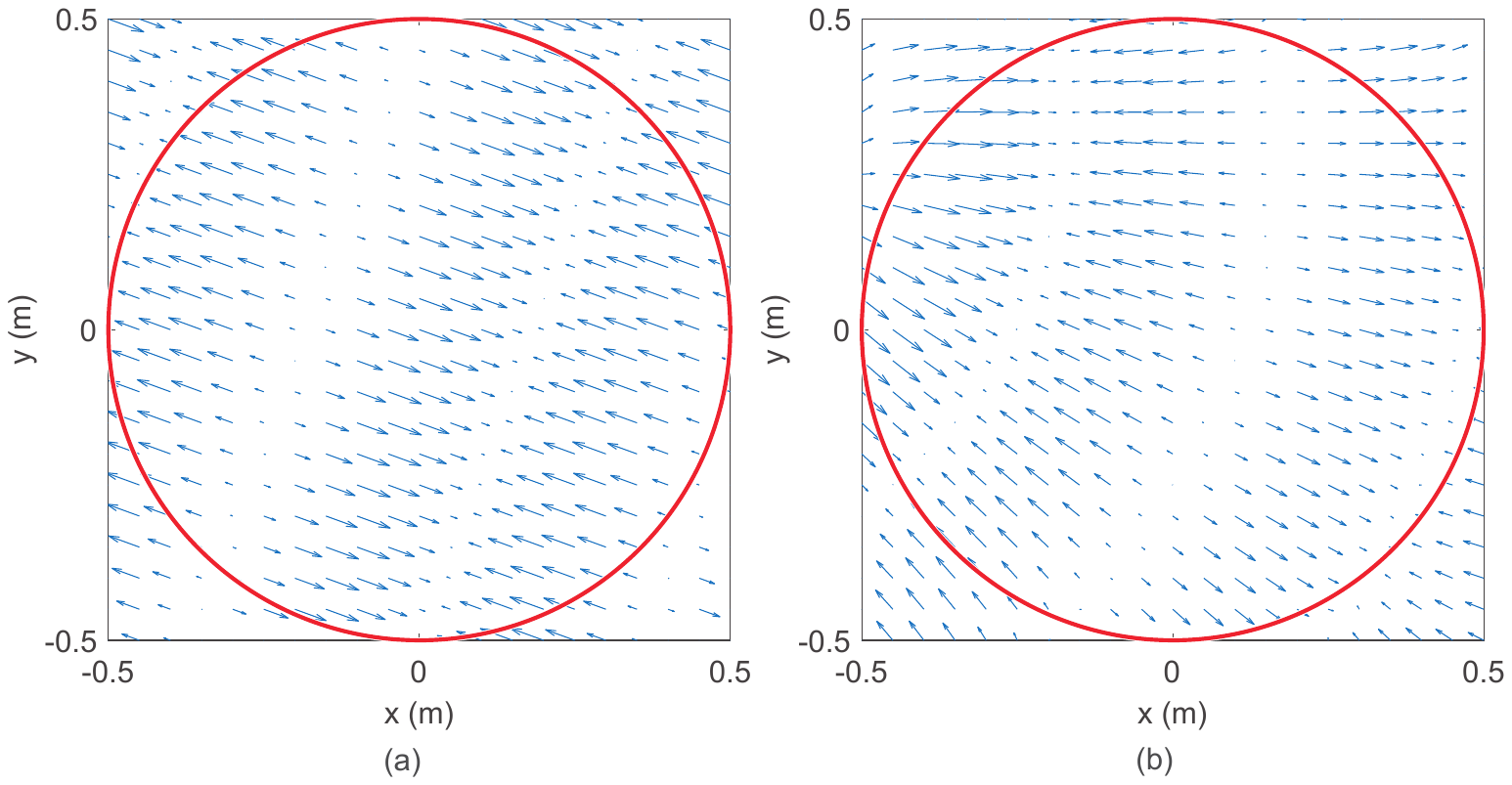}
  \vskip -3mm
  \caption{Real part of the acoustic velocity vectors at 500 Hz. (a) The source is a plane wave with incident direction $\phi_{\text{pw}} = 8\pi/9$ rad. (b) The source is a point source at $\mathbf{r_{s}} = (1 \text{ m}, 8\pi/9 \text{ rad})$.}
  \vskip -3mm
  \label{Fig:velocity_PW_ls}
  \vskip -2mm
\end{figure}

Figure \ref{Fig:velocity_PW_ls} shows the simulation results. The circular area bounded by the red circle has radius 0.5 meters. The cylindrical harmonic coefficients $\beta_{\nu}(k)$ of the pressure are truncated to order $\nu = \pm 7$. Hence, the CHV-indR coefficients $(\zeta_{\boldsymbol{\hat{x}}})_{n}(k)$ are calculated up to order $n = \pm 6$. Figure \ref{Fig:velocity_PW_ls}(a) illustrates the real part of the acoustic velocity vectors when the source is a plane wave with incident direction $\phi_{\text{pw}} = 8\pi/9$ rad (160 degrees). The acoustic velocity vectors are pointing towards either $8\pi/9$ rad or $-\pi/9$ rad. This is because the acoustic velocity vectors of a plane wave are perpendicular to the wave front. Figure 2(b) shows the real part of the acoustic velocity vectors when the source is an infinite line source in $z$ direction located at $\mathbf{r_{s}} = (1 \text{ m}, 8\pi/9 \text{ rad})$ on the two-dimensional $xy$ plane. The acoustic velocity vectors either diverge from or converge to a point in the direction of $8\pi/9$ rad. The cylindrical wave fronts can be discerned. 

\section{Reproducing the acoustic velocity vectors in a circular area}
This section highlights the potential of using the CHV-indR coefficients in sound field reproduction system. In velocity matching (VM), reproduction is achieved by matching the desired CHV-indR coefficients using $L$ sources. The reproduction algorithm consists of the recording stage and the reproduction stage. In the recording stage, first, the cylindrical harmonic coefficients $\beta_{\nu}^{(\text{d})}(k)$ of the pressure of the desired sound field are measured by a circular microphone array. Next, using \eqref{Eq:CHP_2_CHV}, the desired CHV-indR coefficients $(\zeta_{\boldsymbol{\hat{e}}})_{n}^{(\text{d})}(k)$ with $\boldsymbol{\hat{e}} \in \{\boldsymbol{\hat{x}}, \boldsymbol{\hat{y}}\}$ are calculated. In the reproduction stage, the first step is to measure the cylindrical harmonic coefficients $\beta_{\nu}^{(\ell)}(k)$ with $\ell = 1, 2, \cdots, L$ of the pressure in the circular listening area when the input to the $\ell$-th source is a unit sinusoidal signal. Next, using \eqref{Eq:CHP_2_CHV}, the CHV-indR coefficients $(\zeta_{\boldsymbol{\hat{e}}})_{n}^{(\ell)}(k)$ in the circular listening area due to unit sinusoidal input to the $\ell$-th source are found. Then, a system of equations is constructed
\begin{equation}
\label{Eq:CHV_mode_match}
    \boldsymbol{\zeta}^{(\text{d})}(k) = \mathbf{H}(k) \mathbf{w}(k).
\end{equation}
In \eqref{Eq:CHV_mode_match}, $\boldsymbol{\zeta}^{(\text{d})}(k) = [\boldsymbol{\zeta}_{\boldsymbol{\hat{x}}}^{(\text{d})}(k)^{T}, \boldsymbol{\zeta}_{\boldsymbol{\hat{y}}}^{(\text{d})}(k)^{T}]^{T}$ in which the column vector $\boldsymbol{\zeta}_{\boldsymbol{\hat{e}}}^{(\text{d})}(k) $ is the concatenation of $(\zeta_{\boldsymbol{\hat{e}}})_{n}^{(\text{d})}(k)$. The matrix $\mathbf{H}(k) = [\boldsymbol{\zeta}^{(1)}(k), \boldsymbol{\zeta}^{(2)}(k), \cdots, \boldsymbol{\zeta}^{(L)}(k)]$, with its $\ell$-th column $\boldsymbol{\zeta}^{(\ell)}(k) = [\boldsymbol{\zeta}_{\boldsymbol{\hat{x}}}^{(\ell)}(k)^{T}, \boldsymbol{\zeta}_{\boldsymbol{\hat{y}}}^{(\ell)}(k)^{T}]^{T}$ in which the column vector $\boldsymbol{\zeta}_{\boldsymbol{\hat{e}}}^{(\ell)}(k) $ is the concatenation of $(\zeta_{\boldsymbol{\hat{e}}})_{n}^{(\ell)}(k)$. The column vector $\mathbf{w}(k) = [w_{1}(k), w_{2}(k), \cdots, w_{L}(k)]$.
Finally, the weight (driving function) $w_{\ell}(k)$ of each source is found by solving \eqref{Eq:CHV_mode_match}. Suppose the cylindrical harmonic coefficients $\beta_{\nu}^{(\text{d})}(k)$ and $\beta_{\nu}^{(\ell)}(k)$ of the pressure are measured up to $\nu = \pm V$, then the dimension of $\mathbf{H}(k)$
is $[2\times(2V-1)]$-by-$L$. This is because the CHV-indR coefficients $(\zeta_{\boldsymbol{\hat{e}}})_{n}^{(\text{d})}(k)$ and $(\zeta_{\boldsymbol{\hat{e}}})_{n}^{(\ell)}(k)$ can only be calculated up to $n = \pm (V-1)$.

This section compares VM with PM, which is a pressure matching approach that matches the cylindrical harmonic coefficients of the pressure in the listening area. In PM, a system of equation is constructed
\begin{equation}
\label{Eq:CHP_mode_match}
    \pmb{\beta}^{(\text{d})}(k) = \mathbf{G}(k) \mathbf{w}(k).
\end{equation}
In \eqref{Eq:CHP_mode_match}, the column vector $\pmb{\beta}^{(\text{d})}(k)$ is formed by concatenating $\beta_{\nu}^{(\text{d})}(k)$. The matrix $\mathbf{G}(k) = [\pmb{\beta}^{(1)}(k), \pmb{\beta}^{(2)}(k), \cdots, \pmb{\beta}^{(L)}(k)]$ in which the $\ell$-th column $\pmb{\beta}^{(\ell)}(k)$ is the concatenation of $\beta_{\nu}^{(\ell)}(k)$. The dimension of $\mathbf{G}(k)$ is $(2V+1)$-by-$L$.

\begin{figure}[t]
  \centering
  \includegraphics[trim = 30mm 65mm 25mm 70mm, clip, width = 0.9\columnwidth]{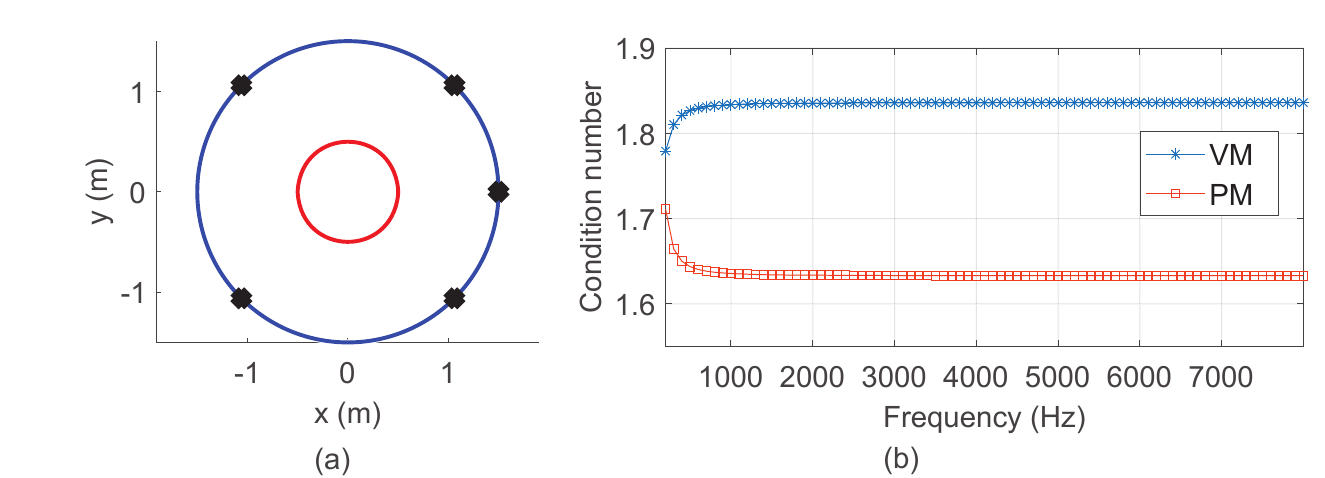}
  \vskip -3mm
  \caption{(a) Setup of the reproduction system. The listening area bounded by the red circle has radius 0.5 meters. Black crosses denote loudspeakers. (b) Condition numbers of $\mathbf{H}(k)$ in VM and $\mathbf{G}(k)$ in PM.}
  \vskip -3mm
  \label{Fig:reprod_setup}
  \vskip -2mm
\end{figure}

Figure \ref{Fig:reprod_setup}(a) shows the simulation setup. The loudspeakers are assumed to be infinite line sources in $z$ direction located on the $xy$ plane. The five loudspeakers are located on the blue circle of radius 1.5 meters and has azimuth angles $[0, \pi/4, 3\pi/4, 5\pi/4, 7\pi/4]$ rad. The circular listening area has radius 0.5 meters and is bounded by the red circle. The desired sound field is a plane wave with incident direction $\phi_{\text{pw}} = 8\pi/9$ rad. The cylindrical harmonic coefficients of the pressure are truncated to $\nu = \pm 3$, which can be measured by a circular microphone array with more than 7 microphones. Hence, the CHV-indR coefficients are truncated to $n = \pm 2$. At each wavennumber $k$, the dimension of $\mathbf{H}(k)$ is 10-by-5 and the dimension of $\mathbf{G}(k)$ is 7-by-5. The condition numbers of $\mathbf{H}(k)$ and $\mathbf{G}(k)$ are shown in Figure \ref{Fig:reprod_setup}(b). The condition numbers are stable, though those of $\mathbf{H}(k)$ exceed those of $\mathbf{G}(k)$. The loudspeaker weights $\mathbf{w}(k)$ in \eqref{Eq:CHV_mode_match} and \eqref{Eq:CHP_mode_match} are found by using the Moore-Penrose pseudoinverse, which is calculated by using the \texttt{pinv} function in MATLAB with default tolerance. A better tolerance value can be incorporated in future work. 

\begin{figure}[t]
  \centering
  \includegraphics[trim = 5mm 50mm 10mm 66mm, clip, width = 0.9\columnwidth]{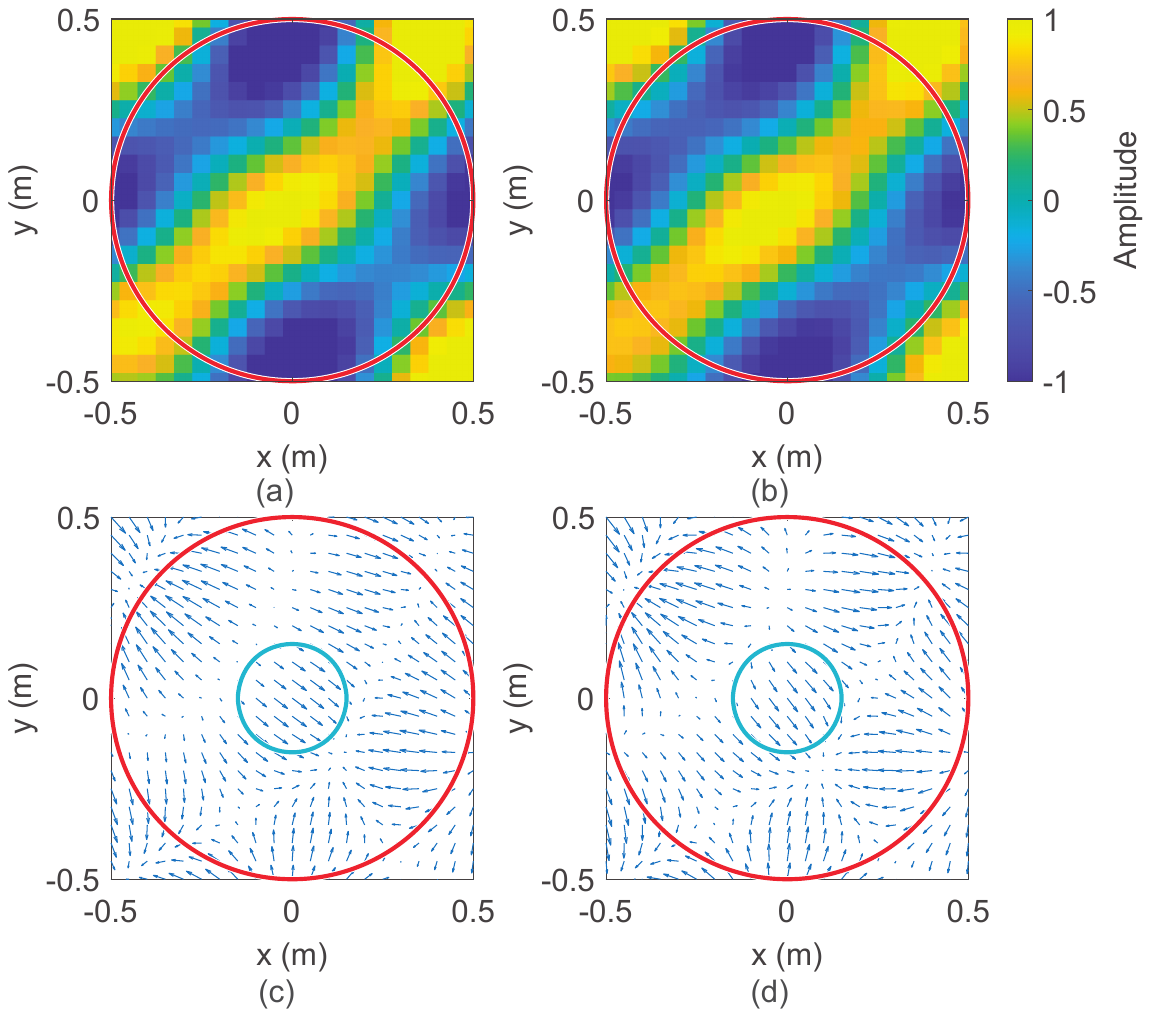}
  \vskip -5mm
  \caption{Real part of the reproduced pressure and the reproduced acoustic velocity vectors at 500 Hz. The desired sound field is a plane wave with incident direction $8\pi/9$ rad. (a) and (c) - VM; (b) and (d) - PM.}
  \label{Fig:reprod_result_160}
  \vskip +2mm
  \centering
  \includegraphics[trim = 35mm 90mm 40mm 95mm, clip, width = 0.9\columnwidth]{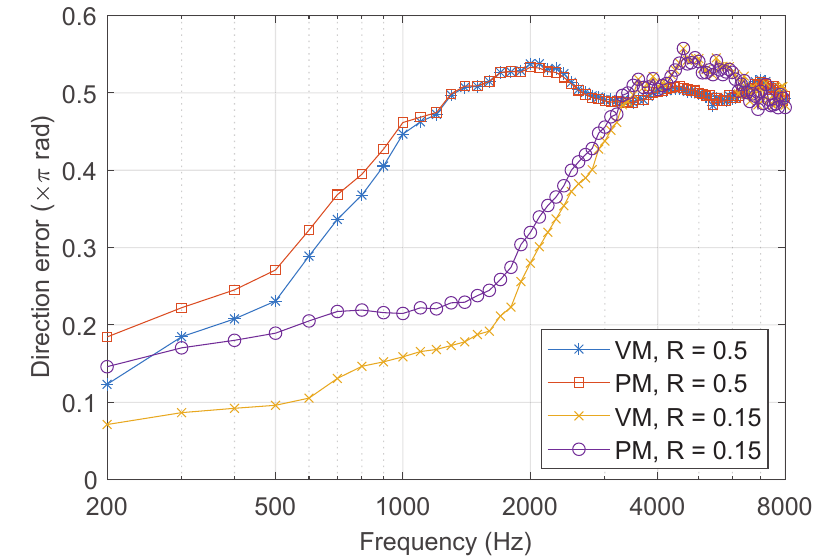}
  \vskip -4mm
  \caption{Direction errors in the real part of the reproduced acoustic velocity vectors. }
  \label{Fig:reprod_error}
\end{figure}

Figure \ref{Fig:reprod_result_160} illustrates the reproduced pressure and the acoustic velocity vectors at 500 Hz. Figures \ref{Fig:reprod_result_160}(a) and \ref{Fig:reprod_result_160}(c) are from VM proposed in this paper, while Figures \ref{Fig:reprod_result_160}(b) and \ref{Fig:reprod_result_160}(d) are from PM. The ground truth of the acoustic velocity vectors is in Figure \ref{Fig:velocity_PW_ls}(a). As in \cite{Lachlan2021} and \cite{ZuoIntensity}, the direction error 
\begin{equation}
    \epsilon(k) = \cos^{-1} (\text{DOT}(k)) \; \text{rad}
\end{equation}
with
\begin{equation}
   \text{DOT}(k) = \frac{\mathbf{V}^{(\text{d})} (\mathbf{r}_{q}, k)}{||\mathbf{V}^{(\text{d})} (\mathbf{r}_{q}, k)||_{2}} \cdot \bigg[\frac{\mathbf{V}^{(\text{r})} (\mathbf{r}_{q}, k)}{||\mathbf{V}^{(\text{r})} (\mathbf{r}_{q}, k)||_{2}} \bigg]^{T}
\end{equation}
in which the desired acoustic velocity vector $\mathbf{V}^{(\text{d})}(\mathbf{r}_{q}, k) \equiv [V_{\boldsymbol{\hat{x}}}^{(\text{d})}(\mathbf{r}_{q}, k), V_{\boldsymbol{\hat{y}}}^{(\text{d})}(\mathbf{r}_{q}, k)]$ and the reproduced acoustic velocity vector  $\mathbf{V}^{(\text{r})} (\mathbf{r}_{q}, k) \equiv [V_{\boldsymbol{\hat{x}}}^{(\text{r})}(\mathbf{r}_{q}, k), V_{\boldsymbol{\hat{y}}}^{(\text{r})}(\mathbf{r}_{q}, k)]$. Figure \ref{Fig:reprod_error} shows the direction errors. The blue line and the red line show the direction errors averaged across 2821 evaluation points within the red circle of radius 0.5 meters in Figure \ref{Fig:reprod_setup}(a). Below 1 kHz, VM achieved lower direction errors than PM. Above 1 kHz, the VM and the PM method have similar direction errors. The yellow line and the purple line illustrate the direction errors averaged across 249 evaluation points within the cyan circle of radius 0.15 meters located at the center of the listening area. VM achieved significant improvement up to 2 kHz. The acoustic velocity vectors is most relevant to human's localization of sound at low frequencies \cite{Gerzon1992, gerzon1992G}. Therefore, this section shows that the VM method could improve localization, when the listener is near the listening area's center. For mid to high frequencies, intensity based reproduction method \cite{ZuoIntensity, ZuoIntensityEU, ZuoIntensityMZ, Choi2004, Arteaga2013} should be considered. Note that the reproduction error will be different when the desired plane wave is coming from a different direction. Future work should also consider the presence of the listener in the reproduced sound field, which could cause performance degradation. Moreover, the selection of the circular microphone array, e.g., the radius and the number of microphones, should be investigated in future work.  

\section{Conclusion}
This paper presented the CHV-indR coefficients, which were the radial independent cylindrical harmonic coefficients of the acoustic velocity vectors in a circular area. By using the sound field translation formula, the CHV-indR coefficients were derived from the cylindrical harmonic coefficients of the pressure inside the circular area. Hence, CHV-indR coefficients could be derived from circular microphone array measurements. The CHV-indR coefficients were used in sound field reproduction system. Simulation showed that compared with method that matches the cylindrical harmonic coefficients of the pressure, the proposed method that matches the CHV-indR coefficients achieved lower direction error in the reproduced acoustic velocity vectors at low frequencies, where acoustic velocity vectors are most relevant to human's localization. 

\bibliographystyle{IEEEtran}
\bibliography{refs23}

\end{document}